\documentclass{article}
\usepackage{spconf,amsmath,graphicx,amssymb,mathrsfs,balance,diagbox,appendix,booktabs}
\usepackage[colorlinks,linkcolor=black]{hyperref}
\usepackage{booktabs}
\usepackage{multirow}

\title{DESNet: A Multi-channel Network for Simultaneous Speech Dereverberation, Enhancement and Separation}

\name{Yihui Fu\textsuperscript{1,*,**},
      Jian Wu\textsuperscript{1,2,*}\thanks{* Contribute equally},
      Yanxin Hu\textsuperscript{1},
      Mengtao Xing\textsuperscript{1},
      Lei Xie\textsuperscript{1,**}\thanks{** Corresponding author}}
\address{\textsuperscript{1}Audio, Speech and Language Processing Group (ASLP), School of Computer Science,
       \\Northwestern Polytechnical University, Xi'an, China
       \\\textsuperscript{2}Speech and Language Group, Microsoft}

\begin{document}

\linespread{0.75}
\maketitle

\begin{abstract}
In this paper, we propose a multi-channel network for simultaneous speech dereverberation, enhancement and separation (DESNet). To enable gradient propagation and joint optimization, we adopt the attentional selection mechanism of the multi-channel features, which is originally proposed in end-to-end \textit{unmixing}, \textit{fixed-beamforming} and \textit{extraction} (E2E-UFE) structure. Furthermore, the novel deep complex convolutional recurrent network (DCCRN) is used as the structure of the speech unmixing and the neural network based weighted prediction error (WPE) is cascaded beforehand for speech dereverberation. We also introduce the \textit{staged SNR} strategy and \textit{symphonic loss} for the training of the network to further improve the final performance. Experiments show that in non-dereverberated case, the proposed DESNet outperforms DCCRN and most state-of-the-art structures in speech enhancement and separation, while in dereverberated scenario, DESNet also shows improvements over the cascaded WPE-DCCRN networks.
\end{abstract}
\section{Introduction}
\label{sec:intro}

Thanks to the rapid development of deep learning in recent years, speech enhancement, separation and dereverberation have received remarkable progress and consistent improvements on public datasets, e.g., WSJ0-2mix~\cite{hershey2016deep}, VoiceBank-DEMAND~\cite{valentini2016investigating} and REVERB Challenge~\cite{kinoshita2016summary}. Various deep neural network (DNN) architectures are proposed and reported to achieve significant improvements on speech enhancement~\cite{tan2018gated,yin2020phasen,hu2020dccrn} or separation~\cite{luo2019conv,luo2020dual,liu2019divide,zeghidour2020wavesplit,takahashi2019recursive} tasks. However, most of the models mentioned above focus on individual enhancement or separation task and haven't considered the real-world environment that speech overlapping, directional/isotropic noise and reverberation may exist together, which leads us to consider adopting one universal model to cope with speech enhancement, separation and dereverberation simultaneously.

Several works have been done before on joint speech enhancement and separation. The most straightforward way is to train a separation model directly using noisy mixtures~\cite{wichern2019wham}. In general, the model can learn to recover each denoised speaker in the mixtures without any modification on the training objective and pipeline. Another popular way is to form a cascaded model or two-stage system following the \textit{enhancement-separation}~\cite{maciejewski2020whamr,ma2020two} or \textit{separation-enhancement}~\cite{wang2018integrating,yoshioka2019low} scheme. In \cite{wu2020end}, an end-to-end structure following the \textit{separation-enhancement} processing is proposed, which enables the joint optimization of the speech unmixing and extraction and yield impressive improvement in online scenario. Besides, the work of \cite{wu2020saddel} is proposed to increase the robustness of noisy speech separation under the recursive separation framework~\cite{takahashi2019recursive,kinoshita2018listening} and studies in ~\cite{von2019all,kinoshita2020tackling} continue to consider the practical issue of recursive approach in real meeting scenarios. The authors of  \cite{nakatani2020dnn} investigate the DNN supported system that integrates conventional spatial-clustering and beamforming techniques.

\begin{figure*}[htb]
\centering
\includegraphics[width=1 \textwidth]{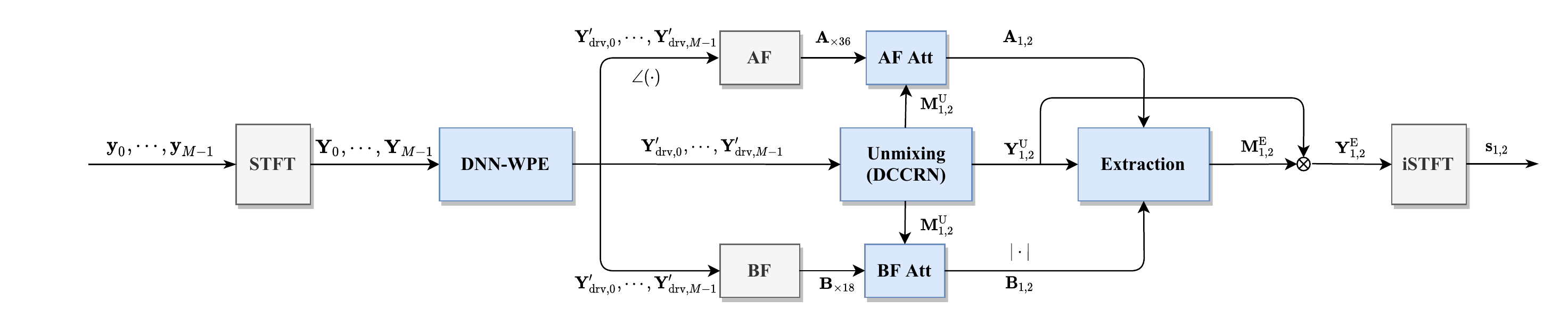}
  \vspace{-0.7cm}
  \caption{Overall framework of proposed DESNet.}
\label{fig:DESNet}
\end{figure*}

In this paper, we focus on offline processing in far-field multi-channel scenario and propose an all neural multi-channel network for simultaneous speech dereverberation, enhancement and separation named \textit{DESNet}. It utilizes the DNN based weighted prediction error (WPE)~\cite{kinoshita2017neural} as the dereverberation module, which estimates the time-varying variance of the early reverberated signal using DNN instead of the iterative procedure in original WPE. The design of the following \textit{separation-enhancement} processing is motivated by the E2E-UFE network~\cite{wu2020end}, which consists of three components, namely speech unmixing, attentional feature selection and speech extraction. Different from the original E2E-UFE, the outputs of the speech unmixing network, together with the weighted beams and angle features are concatenated as the input of the speech extraction while the weighted beams and angle features here are treated as the assisted multi-channel directional feature. A similar usage can be found in \cite{wang2018integrating}, where the log spectra of the enhanced signal after the multi-channel Wiener filter is adopted. We use recently proposed deep complex convolutional recurrent network~\cite{hu2020dccrn} (DCCRN) as the structure of speech unmixing. This advanced network achieves the best MOS in real-time track and the second-best in non-real-time track according to the P.808 subjective evaluation in 2020 Deep Noise Suppression (DNS) challenge~\cite{reddy2020interspeech}. The time domain signal is synthesized via inverse short-time Fourier transform (iSTFT) using the final enhanced or separated spectra from the speech extraction module. The DNN-WPE, speech unmixing, attentional feature selection and speech extraction have no non-differentiable operations. Thus we can optimize DESNet jointly in an end-to-end manner. The overall framework of DESNet is shown in Fig.~\ref{fig:DESNet}.

We evaluate the performance of the proposed model on three tracks: speech enhancement (SE), clean speech separation (CSS) and noisy speech separation (NSS), in two categories: dereverberated and non-dereverberated. In order to improve the overall performance in each track,  we introduce the \textit{staged SNR} strategy and \textit{symphonic loss} for the network training. Experiments show that in non-dereverberated scenario, the proposed DESNet outperforms DCCRN in enhancement track and most state-of-the-art structures in separation task, e.g., DPRNN~\cite{luo2020dual}, FasNet~\cite{Luo2019End}, while in dereverberated scenario, DESNet also shows improvements over the cascaded WPE-DCCRN network in all those three tracks.

The rest of the paper is organized as follows. In Section 2, we give a detailed introduction of DESNet. In Section 3, the experimental settings, training details, evaluation scheme, results and analysis will be discussed. A conclusion of our work is drawn in Section 4.

\section{Proposed DESNet}

\subsection{Problem Formulation}

We consider the far-field signal model with $C$ speakers in time-domain as follow:
\begin{equation}
    \mathbf{y}_m  = \sum_{c} \mathbf{x}_{c,m} + \mathbf{n}_{m} = \sum_{c} \mathbf{s}_c * \mathbf{h}_{c,m} + \mathbf{n}_{m},
\end{equation}
where $\mathbf{x}_{c,m}$ denotes the image of the $c$-th speaker ($0 \leqslant c < C$) at the position of the $m$-th microphone ($0 \leqslant m < M$). $\mathbf{h}_{c,m}$ is room impulse response (RIR) between speaker $c$ and microphone $m$. $\mathbf{s}_c$ is $c$-th source speaker signal and we model environment noise at microphone $m$ as $\mathbf{n}_m$. The corresponding frequency domain signal model is shown as:
\begin{equation}
        \mathbf{Y}_m  = \sum_{c} \mathbf{X}_{c,m}  + \mathbf{N}_{m},
\end{equation}
where $\{\mathbf{Y}_m, \mathbf{X}_{c,m}, \mathbf{N}_{m}\} \in \mathbb{C}^{T \times F}$ are short-time Fourier transform (STFT) of $\{\mathbf{y}_m, \mathbf{x}_{c,m}, \mathbf{n}_{m}\}$. $T$ and $F$ denote the total number of time frames and frequency bins.

\subsection{Dereverberation}
WPE is an algorithm to effectively reduce reverberation and greatly improve the speech quality. Given a group of estimated linear filter weights $\mathbf{G}_f \in [\mathbf{G}_{0,f}, \cdots, \mathbf{G}_{K-1,f}]^\top \in \mathbb{C}^{MK \times M}$, where $K$ denotes tab number, the dereverberated signal is obtained via:
\begin{align}
    \mathbf{y}'_{\text{drv},t,f} & = \mathbf{y}_{t,f} - \sum_{k=0}^{K -1} \mathbf{G}_{f,k}^H \mathbf{y}_{t - \Delta -k,f} \\
    & = \mathbf{y}_{t,f} - \mathbf{G}_{f}^H \overline{\mathbf{y}_{t - \Delta,f}},
\end{align}
where $\Delta$ denotes prediction delay, $\mathbf{y}_{t,f} = [\mathbf{Y}_{0,t,f}, \cdots, \mathbf{Y}_{M - 1,t,f}]^T \\ \in \mathbb{C}^{M \times 1}$ and $\overline{\mathbf{y}_{t - \Delta,f}} = [\mathbf{y}_{t - \Delta,f}^T, \cdots, \mathbf{y}_{t - \Delta - K + 1,f}^T]^T \in \mathbb{C}^{MK \times 1}$.

The estimation of $\mathbf{G}_f$ is based on the time-varying variance $\mathbf{\Lambda} \in \mathbb{R}^{T \times F}$ of the early reverberated signal following the steps:
\begin{align}
    \mathbf{R}_f &= \sum_t \frac{\overline{\mathbf{y}_{t - \Delta,f}} \overline{\mathbf{y}_{t - \Delta,f}}^H}{\mathbf{\Lambda}_{t,f}}, \\
    \mathbf{r}_f &= \sum_t \frac{\overline{\mathbf{y}_{t - \Delta,f}} \mathbf{y}_{t,f}^H}{\mathbf{\Lambda}_{t,f}}, \\
    \mathbf{G}_f &= \mathbf{R}_f^{-1} \mathbf{r}_f.
\end{align}

The conventional WPE estimates $\mathbf{\Lambda}$ in an iterative procedure, while in DNN-WPE~\cite{kinoshita2017neural}, $\mathbf{\Lambda}$ is estimated using a neural network directly:
\begin{align}
    \mathbf{\Lambda}_m & = \text{NN}(|\mathbf{Y}_m|),
\end{align}
and $\mathbf{\Lambda} = \sum_m \mathbf{\Lambda}_m / M$.

\subsection{Angle Feature and Fixed Beamforming}
After estimating the dereverberated signal on frequency $f$ $\mathbf{Y}'_{\text{drv},f} = [\mathbf{y}'_{\text{drv},0,f}, \cdots, \mathbf{y}'_{\text{drv},T - 1,f}] \in \mathbb{C}^{M \times T}$, we calculate angle features (in subnet AF) and fixed beams (in subnet BF) in all candidate directions. The beamformed signal $\mathbf{b}_{i,f} \in \mathbb{C}^{1 \times T}$ on $i$-th ($0 \leqslant  i < N_b$) direction is obtained via:
\begin{equation}
\mathbf{b}_{i,f} = \mathbf{w}_{i,f}^{H} \mathbf{Y}'_{\text{drv},f},
\end{equation}
where $\mathbf{w}_{i,f} \in \mathbb{C}^{M \times 1}$ represents the $i$-th beam weight on frequency $f$. We denote $\mathbf{B}_i = [\mathbf{b}_{i,0}^T, \cdots, \mathbf{b}_{i,F - 1}^T]^T \in \mathbb{C}^{T \times F}$ in this paper.

The angle feature $\mathbf{a}_{\theta,f} \in \mathbb{R}^{1 \times T}$ on frequency band $f$ is derived from:
\begin{equation}
\mathbf{a}_{\theta,f} = \sum_{m, n \in \psi } \cos (\mathbf{o}_{mn, f} - \mathbf{r}_{\theta,mn, f}) / P,
\end{equation}
where $\psi$ contains $P$ microphone pairs and $\mathbf{o}_{mn, f} = \angle \mathbf{y}'_{\text{drv},m, f} \\ - \angle \mathbf{y}'_{\text{drv},n, f}$ represents the observed inter-microphone phase difference (IPD) between microphone $m$ and $n$. The reference IPD $\mathbf{r}_{\theta, mn, f}$ can derive from the geometry-dependent steer vector given the DoA $\theta$ and microphone index $m$ and $n$. Similar with $\mathbf{B}_i$, we define $\mathbf{A}_\theta = [\mathbf{a}_{\theta,0}^T, \cdots, \mathbf{a}_{\theta, F - 1}^T]^T \in \mathbb{R}^{T \times F}$. The total number of candidate DoA is defined as $N_{A}$.
\subsection{Speech Unmixing by DCCRN}
\begin{figure}[htb]
\centering
  \centerline{\includegraphics[width=0.35 \textwidth]{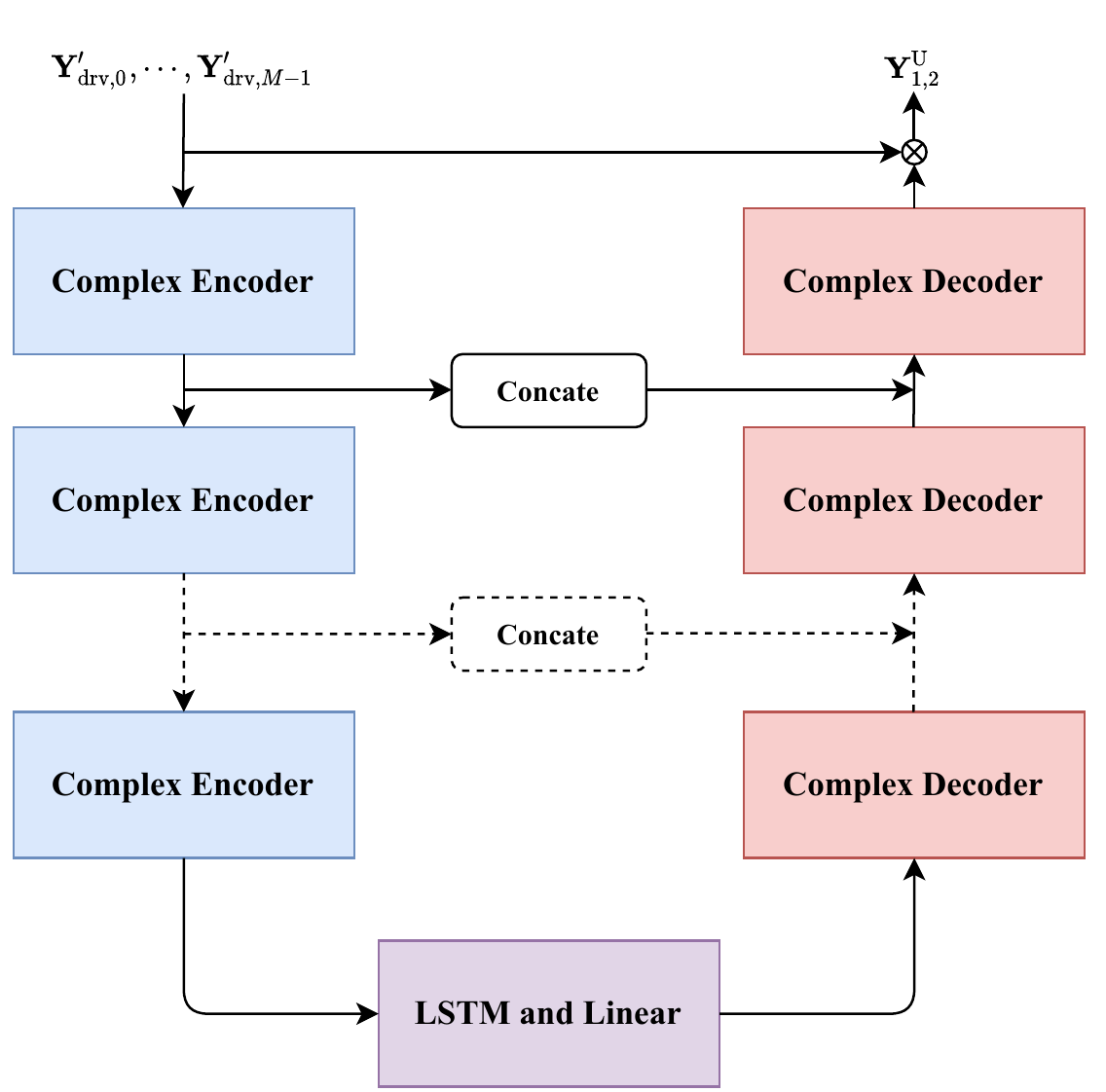}}
  \caption{Diagram of DCCRN.}
  \vspace{-0.5cm}
\label{fig:DCCRN}

\end{figure}

\begin{figure}[htb]
  \centering
  \centerline{\includegraphics[width=4.5cm]{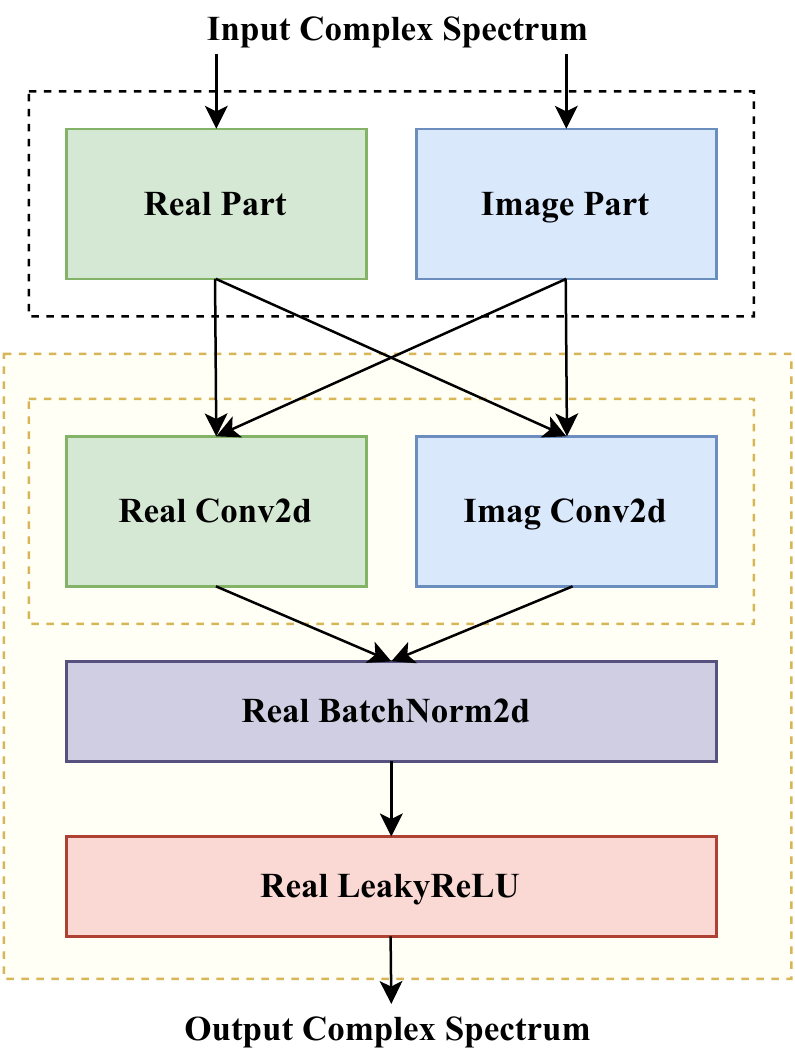}}
  \vspace{-0.1cm}
\caption{Diagram of complex-valued convolutional encoder/decoder.}
\label{fig:complex}
\end{figure}

Compared with the original E2E-UFE whose unmixing module only consists of stacked LSTMs and a projection layer,  we believe that a better network prototype, like DCCRN~\cite{hu2020dccrn}, can benefit the following selection of the angle and beam features, as well as assist the speech extraction for a better estimation of the final masks. DCCRN follows the conventional UNet structure, but using complex-valued convolutional encoders/decoders and real/imaginary LSTMs to model the context dependency.  The architecture of DCCRN is shown in Fig.~\ref{fig:DCCRN}, where the encoder/decoder block is stacked by 2D complexed convolution/deconvolution layers, real-valued 2D batch normalization (BN) and leaky ReLU activation function, as shown in Fig.~\ref{fig:complex}. Similar to the implementation in DCUNet, the complex-valued 2D convolutional operation is interpreted as two real-valued ones, following the formula:
\begin{equation}
\mathbf{W} \circledast \mathbf{Y} =
\begin{bmatrix}
\mathbf{W}_r \\
\mathbf{W}_i
\end{bmatrix} \circledast
\begin{bmatrix}
\mathbf{Y}_r \\
\mathbf{Y}_i
\end{bmatrix}
=
\begin{bmatrix}
\mathbf{W}_r * \mathbf{Y}_r - \mathbf{W}_i * \mathbf{Y}_i \\
\mathbf{W}_r * \mathbf{Y}_i + \mathbf{W}_i * \mathbf{Y}_r
\end{bmatrix},
\end{equation}
where $\circledast$ and $*$ denote complex-valued and real-valued convolution, respectively. $\mathbf{W} = [\mathbf{W}_r, \mathbf{W}_i]^T$ is complex-valued convolution filters and  $\mathbf{Y} = [\mathbf{Y}_r, \mathbf{Y}_i]^T$ is the input of complexed convolution layer. As DCCRN achieves impressive performance on speech enhancement task, it inspires us to adopt it and extent its ability to separation task. We use real valued LSTM with a linear projection layer to map the concatenation of real and image part of the encoder output to two branches and the decoder is responsible for the generation of the pre-unmixing masks for each speaker independently. Structure of the decoders is symmetrical with the encoders, replacing 2D convolution with 2D deconvolution. The output of each encoder layer is concatenated to the output of the corresponding decoder layer to avoid gradient vanishing.

The final complex ratio mask (CRM) of speaker-$c$ $\mathbf{M}_c^U = \mathbf{M}_{c,r} + i\mathbf{M}_{c,i}$ is calculated given the output of the decoders $\mathbf{H}_c = [\mathbf{H}_{c,r}, \mathbf{H}_{c,i}]^T$:
\begin{align}
\mathbf{M}_{c,\text{mag}} & =\tanh(\sqrt{\mathbf{H}_{c,r}^2+\mathbf{H}_{c,i}^2}), \\
\mathbf{M}_{c,\text{pha}} & =\text{arctan2}(\mathbf{H}_{c,i}, \mathbf{H}_{c,r}).
\end{align}
And then we have $\mathbf{M}_{c,r} = \mathbf{M}_{c,\text{mag}} \odot \cos(\mathbf{M}_{c,\text{pha}}), \mathbf{M}_{c,i} = \mathbf{M}_{c,\text{mag}} \odot \sin(\mathbf{M}_{c,\text{pha}})$.
Thus we can get unmixed results $\mathbf{Y}^U_c \in \mathbb{C}^{T \times F}$ using the dereverberated speech at channel 0:
\begin{equation}
    \mathbf{Y}^U_c = \mathbf{M}^U_c \odot \mathbf{Y}'_{\text{drv}, 0}.
\end{equation}

\subsection{Attentional Feature Selection}
After figuring out all candidate $N_{A}$ angles and $N_{B}$ beams from subnet AF and BF as well as unmixing mask $\mathbf{M}_c^{U} \in \mathbb{C}^{T \times F}, c \in \{1,2\}$, we utilize the proposed attentional selection mechanism from \cite{wu2020end} for multi-channel feature selection. Here we take angle feature as example and similar progress is applied to obtain beam feature. First, two sets of embedding for unmixing masks $\mathbf{M}_c^{U}$ of each speaker and angle features $\mathbf{A}_{\theta}$ are formed through linear mapping:
\begin{align}
\mathbf{V}_{c}^{U} &= |\mathbf{M}_{c}^{U}|\mathbf{W}_{p}, \\
\mathbf{V}_{\theta}^{A} &= \mathbf{A}_{\theta}\mathbf{W}_{a},
\end{align}
where $\mathbf{W}_{p}, \mathbf{W}_{a} \in \mathbb{R}^{F\times D}$ are linear transform weights, resulting in the final embedding matrices $\mathbf{V}_{c}^{U}, \mathbf{V}_{\theta}^{A} \in \mathbb{C}^{T \times D}$. Then a pair-wised similarity matrix is derived between each
frame in $\mathbf{V}_c^U$ and $\mathbf{V}_{\theta}^A$ using dot product distance,
scaled by $(\sqrt{D})^{-1}$:
\begin{align}
    s_{c,\theta,t} &= (\sqrt{D})^{-1} \left(\mathbf{V}^U_{c,t} \right)^T \mathbf{V}^A_{\theta,t}.
    \label{eq7}
\end{align}
The similarity matrix is then averaged along time axis, followed by a softmax activation, to generate the weight $w_{c,\theta}$ of each direction for each source speaker:
\begin{align}
    \hat{s}_{c,\theta} & = \left( T \right)^{-1} \sum_t s_{c,\theta,t}, \\
    w_{c, \theta} &= \text{softmax}_\theta(\hat{s}_{c,\theta}).
\end{align}
Finally, the weight average operation is performed in order to get the weighted angle feature $\hat{\mathbf{A}}_c$ for $c$-th speaker, as shown in the following:
\begin{equation}
\hat{\mathbf{A}}_{c}=\sum_{\theta} w_{c,\theta}\mathbf{A}_{\theta}. \label{eq10}
\end{equation}

\subsection{Speech Extraction}
After generating the unmixed speech from DCCRN as well as the attentional angle and beam features from the AF Att and BF Att layers, we concatenate these three features and feed them to the following speech extraction network. The extractor consists of a stack of LSTM layers, followed by a linear projection to estimate the final real masks $\mathbf{M}_c^{E} \in \mathbb{R}^{T \times F}$. We apply it to the real and imaginary parts of the unmixed speech of each speaker separately to generate the final separation or enhancement outputs.

\subsection{Loss Function}
The proposed model works in an end-to-end manner which directly operates on raw waveform and generates the enhanced/separated results. The clean speech from the primary channel (channel 0 in this paper) is adopted as the training target. Scale-invariant source-to-noise ratio (SI-SNR)~\cite{le2019sdr} is used as the objective function, which has been widely used to replace the standard source-to-distortion ratio (SDR) or mean square error (MSE) loss function. SI-SNR is calculated by:
\begin{equation}
\text{SI-SNR}(\mathbf{s}_i, \mathbf{x}_j) = 20\log_{10}\frac{\Vert \alpha \cdot \mathbf{x}_j \Vert}{\Vert \mathbf{s}_i -\alpha \cdot \mathbf{x}_j \Vert},
\end{equation}
where $\mathbf{x}_j$ is the clean reference of speaker $j$, and $\mathbf{s}_i$ refers to the separated signal of speaker $i$. $\alpha$ is a optimal scaling factor computed via $\alpha  = \mathbf{s}_{i}^\text{T}\mathbf{x}_{j}/\mathbf{x}_{j}^\text{T}\mathbf{x}_{j}$.

In particular, we propose a special way to compute the training loss named symphonic loss for optimization of the DESNet. The loss calculation of each training chunk in one mini-batch is different depending on which track it belongs to. If current mixture chunk contains one speaker, namely in SE track, we only optimize the first branch of the network, while for NSS and CSS tracks, we optimize both branches of the network using permutation invariant training (PIT) strategy, which can be written as:
\begin{equation}
    \mathcal{L} =  -\max_{\phi \in \mathcal{P}} \sum_{(i, j) \in \phi} \text{Si-SNR}(\mathbf{s}_i, \mathbf{x}_j) / N_{\mathcal{P}},
    \label{pit}
\end{equation}
where $\mathcal{P}$ and $N_{\mathcal{P}}$ refer to all possible permutations and permutation number, respectively.

\section{Experiments}
\label{sec:pagestyle}
\subsection{Datasets}
In our experiments, we adopt \textit{train-clean-100} and \textit{train-clean-360} in LibriSpeech which include 460 hours single-channel speech as the source data. The noise set provided by DNS Challenge~\cite{reddy2020interspeech} that contains 180 hours data is used as the noise dataset. The multi-channel RIRs and isotropic noise are simulated in advance based on a circular array with four microphones put around. The radius of the circular array is 5 cm. In SE and CSS tracks, the SNR and SDR during training are sampled randomly from [-5, 10] dB and [-5, 5] dB, respectively. In NSS track, the SNR range is set as [5, 20] dB and SDR range is [-5, 5] dB. The maximum number of speaker is set as 2 for the separation task. All training utterances are added with isotropic noise with SNR ranges from 15 dB to 20 dB. The RT60 of RIRs ranges from 0.1 s to 0.5 s. The angle between each speaker and noise is set at least $20^{\circ}$ to ensure the spatial distinctiveness of each sound source.

For model evaluation, we also create three test sets named SE, CSS and NSS in dereverberated and non-dereverberated categories. All speech data is derived from \textit{test-clean} in LibriSpeech while noise data is selected from DNS noise set which has no overlap with training data. The SDR for two speakers in both CSS and NSS are set to $\{-5, -2, 0\}$ dB, while SNR of noise in NSS is sampled randomly in range of [5, 20] dB. Noise SNR in SE is set to $\{-5, -0, 5, 10\}$ dB. Both CSS and NSS contain 3900 utterances and SE contains 5860 utterances in total. Other settings are kept the same as the training configurations.

\begin{table}[]
\centering
\caption{SNR (SDR) range (dB) in each stage.}
\label{table:SNR range}
\begin{tabular}{ccccc}
\toprule
\multirow{2}{*}{Training Epoch} & \multirow{2}{*}{SE} & \multirow{2}{*}{CSS} & \multicolumn{2}{c}{NSS} \\
& & & SE & SS \\
\midrule
1 $\sim$ 5 & [5, 10]  & [-2, 2]  & $\times$ & $\times$ \\
6 $\sim$ 10 & [0, 10]  & $\times$ & [15, 20] & [-2, 2] \\
11 $\sim$ 15 & [-2, 10] & $\times$ & [10, 20] & [-4, 4] \\
16 $\sim$ 20 & [-5, 10] & $\times$ & [5, 20] & [-5, 5] \\
\bottomrule
\end{tabular}
\end{table}

\subsection{Training Setups}
The training utterances are generated on-the-fly and segmented into 4s chunks in one batch. In other words, we randomly select speech, noise and RIR and simulate the mixture utterances dynamically with one or two speakers according to SNR (SDR) ranges on each track to form 265 hours of training data in each epoch. Compared with generating all the data beforehand, our dynamic simulating strategy can improve the diversity of the training samples. Each model is trained for 20 epochs with Adam optimizer. The learning rate is set to 0.001 initially and decays by 0.5 if the validation loss goes up. Furthermore, we propose the staged SNR strategy to gradually optimize the network in order to adapt to the low SNR (SDR) scenario. The SNR (SDR) range in each training stage is shown in Table~\ref{table:SNR range}. Before epoch 6 we only optimize the network with SE and CSS data with higher SNR (SDR). And after epoch 5, we replace CSS data with NSS data and gradually expand the SNR (SDR) range. Both tracks share the same amount of data in one single epoch.

\begin{table*}[!htb]
\centering
\caption{Results of non-dereverberated SE and SS.}
\setlength{\tabcolsep}{6pt}
\label{table:SE+SS}
\begin{tabular}{l ccccc cccc cccc}
\toprule
Model & \multicolumn{5}{c}{SE (PESQ)} & \multicolumn{4}{c}{CSS (SI-SNR (dB))} & \multicolumn{4}{c}{NSS (SI-SNR (dB))} \\
SNR (SDR) & -5 & 0 & 5 & 10 & Avg. & -5 & -2 & 0 & Avg.  & -5 & -2 & 0 & Avg.       \\ \midrule
Mixed & 1.51 & 1.87 & 2.22 & 2.57 & 2.04 & 0.00 & 0.00 & 0.00 & 0.00 & -1.63 & -0.88 & -0.76 & -1.09 \\
CACGMM & 2.14 & 2.40 & 2.69 & 2.88 & 2.53 & 4.50 & 6.16 & 6.48 & 5.71 & 1.72 & 4.08 & 4.46 & 3.42 \\
Proposed DESNet  &\textbf{2.55} & \textbf{2.87} & \textbf{3.17} & \textbf{3.41} & \textbf{3.00} & \textbf{10.18} & \textbf{9.98} & \textbf{9.78} & \textbf{9.98} & \textbf{7.16} & \textbf{7.73} & \textbf{7.77} & \textbf{7.55} \\
$\quad$ - Staged SNR & 2.51 & \textbf{2.87} & 3.16 & 3.40 & 2.99 & 9.88 & 8.54 & 7.87 & 8.76 & \textbf{7.16} & 6.65 & 6.19 & 6.67 \\
$\quad$ - Symphonic Loss & 2.36 & 2.73 & 3.06 & 3.33 & 2.87 & 9.61 & 9.40 & 9.26 & 9.42 & 6.70 & 7.31 & 7.31 & 7.11 \\
$\quad$ - BF Feature & 2.29 & 2.65 & 2.97 & 3.23 & 2.79 & 8.77 & 8.65 & 8.44 & 8.62 & 5.84 & 6.32 & 6.31 & 6.16 \\
DCCRN & 2.25 & 2.61 & 2.94 & 3.20 & 2.75 & 7.78 & 6.04 & 5.37 & 6.40 & 5.73 & 4.62 & 4.07 & 4.81 \\
Conv-TasNet &2.00 &2.29 &2.53 &2.71 &2.38 &6.03 &6.67 &6.72 &6.47 &3.93 &5.09 &5.23 &4.75 \\
DPRNN &2.22 &2.55 &2.84 &3.09 &2.68 &9.09 &9.36 &9.32 &9.26 &6.37 &7.32 &7.42 &7.04 \\
FasNet & 2.24 & 2.58 & 2.89 & 3.14 & 2.71 & 9.42 & 9.35 & 9.02 & 9.26 & 6.91 & 7.63 & 7.41 & 7.32 \\ \bottomrule \hline
\end{tabular}
\vspace{-0.5cm}
\end{table*}

\begin{table*}[!htb]
\centering
\caption{Results of dereverberated SE and SS.}
\setlength{\tabcolsep}{6pt}
\label{table:SE+SS+DRV}
\begin{tabular}{l ccccc cccc cccc}
\toprule
Model & \multicolumn{5}{c}{SE (PESQ)} & \multicolumn{4}{c}{CSS (SI-SNR (dB))} & \multicolumn{4}{c}{NSS (SI-SNR (dB))} \\
SNR (SDR) & -5 & 0 & 5 & 10 & Avg. & -5 & -2 & 0 & Avg.  & -5 & -2 & 0 & Avg.       \\ \midrule
Mixed & 1.41 & 1.71 & 2.02 & 2.31 & 1.86 & -1.38 & -0.75 & -0.64 & -0.92 & -2.63 & -1.54 & -1.35 & -1.84\\
CACGMM & 2.09 & 2.36 & 2.63 & 2.83 & 2.48 & 3.97 & 5.54 & 5.85 & 5.12 & 1.57 & 3.90 & 4.27 & 3.25 \\
Proposed DESNet &\textbf{2.36} &\textbf{2.65} &\textbf{2.90} &\textbf{3.12} &\textbf{2.76} &\textbf{8.07} &\textbf{8.18} &\textbf{8.14} &\textbf{8.13} &\textbf{6.38} &\textbf{6.65} &\textbf{6.50} &\textbf{6.51}\\
$\quad$ - Staged SNR &2.26 &2.57 &2.84 &3.06 &2.68 &7.96 &8.14 &8.03 &8.04 &5.56 &6.36 &6.18 &6.03\\
$\quad$ - Symphonic Loss &2.32 &2.63 &2.89 &3.11 &2.74 &7.74 &7.88 &7.42 &7.68 &5.68 &6.45 &6.50 &6.21 \\
$\quad$ - DNN-WPE &2.17 &2.49 &2.77 &3.01 &2.61 &7.36 &7.66 &7.59 &7.54 &5.20 &5.68 &5.65 &5.51 \\
WPE-DCCRN &2.16 &2.49 &2.78 &3.00 &2.61 &6.64 &6.09 &5.77 &6.17 &5.16 &5.07 &4.61 &4.95  \\
\bottomrule
\end{tabular}
\end{table*}
\subsection{Experiment Setups}
For (i)STFT layer in the DESNet, we use hanning window with a FFT size of 512 and a hop size of 256. $N_A$ and $N_B$ are chosen as 36 and 18 in our experiments. The angle feature calculated in AF layer is averaged among three microphone pairs: (0, 1), (0, 2) and (1, 3). The embedding size $D$ used for feature selection is set as 257. We use 6-layer encoder and decoder in DCCRN with output channel $\{16, 32, 64, 128, 256, 256\}$ and the kernel and stride size are set to (5, 2) and (2, 1), respectively. The hidden size of the 3-layer LSTM in DCCRN is set to 512 and the dimension of the following linear layer is 1024. Speech extraction is a stack of 3-layer LSTM with hidden size of 512.

\subsubsection{Non-dereverberated SE and SS}
In this category, we only perform simultaneous enhancement and separation thus the DNN-WPE is removed from the DESNet. The clean reverberated speech is used as the training reference. CACGMM~\cite{ito2016complex}, DCCRN~\cite{hu2020dccrn}, Conv-TasNet~\cite{luo2019conv}, DPRNN~\cite{luo2020dual} and FasNet~\cite{Luo2019End} are used as the comparative systems.

CACGMM\footnote{\url{https://github.com/funcwj/setk/}} is a blind mask estimation method based on spatial-clustering proposed for speech separation. In this paper, the number of the Gaussian is chosen as 2 for SE and CSS tasks and 3 for NSS task. EM steps are repeated for 50 epochs to estimate the mask of each sound source.

The setups of the multi-channel Conv-TasNet\footnote{\url{{https://github.com/funcwj/conv-tasnet}}} and DPRNN follow the best configurations in \cite{luo2019conv,luo2020dual} but changing input channel size of the encoder from 1 to 4. The window size is chosen as 20 and 16 samples, respectively. The structure of DCCRN is same as the one used in the proposed DESNet. For FasNet\footnote{\url{{https://github.com/yluo42/TAC}}}, 4-channel setups with TAC module are used.

To better validate the proposed structure and strategies, we add three ablation experiments. The first one is to use the fixed SNR range in the last line of the Table~\ref{table:SNR range} to train the model directly. Secondly, we remove the fixed beamformer layer to see whether this learnt multi-channel feature is beneficial to the final performance. We also disable the proposed symphonic loss during training to verify the improvements it brings. The experimental results are shown in Table~\ref{table:SE+SS} and will be discussed in Section 3.4.

\subsubsection{Dereverberated SE and SS}
In this category, we perform simultaneous dereverberation, enhancement and separation. The early reverberation part of the clean source image is used as the training labels. As the time domain separation model used in Section 3.3.1 does not consider dereverberation, we only use CACGMM and WPE-DCCRN as the baseline systems.

The structures of the DNN-WPE network in both DESNet and WPE-DCCRN are the same and it contains a 2-layer CNN whose channel and kernel size are set to (300, 4) and 3, respectively. Other setups are identical to the non-dereverberated category. WPE-DCCRN is also a cascaded network where the input of the DCCRN is the STFT of the early reverberation part of the mixture signal produced by the DNN-WPE module. We also add three ablation experiments to verify the benefits of the staged SNR strategy, symphonic loss and DNN-WPE module, as shown in Table. \ref{table:SE+SS+DRV}.

\begin{figure}[!tbh]
\begin{minipage}[b]{0.50\linewidth}
  \centering
  \centerline{\includegraphics[width=5.0cm]{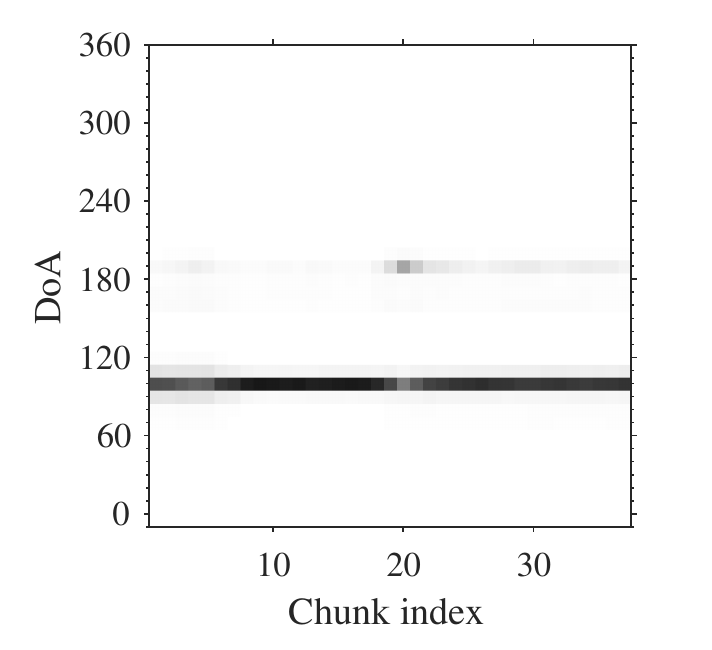}}
\end{minipage}
\hfill
\begin{minipage}[b]{0.40\linewidth}
  \centering
  \centerline{\includegraphics[width=5.0cm]{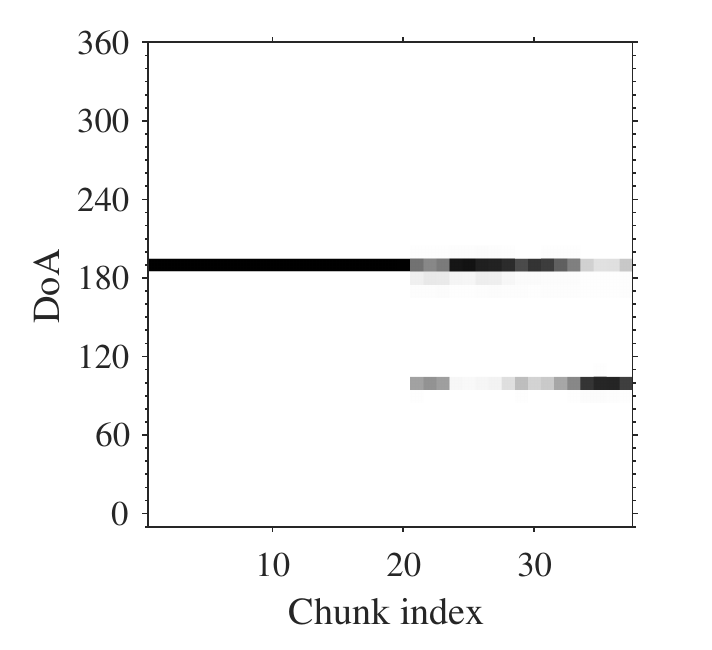}}
\end{minipage}
\vspace{-0.3cm}
\caption{Example of the learnt weights on angle feature in a two-speaker mixture utterance.}
\label{fig:weight}
\end{figure}

\subsection{Results and Discussion}

The experimental results of non-dereverberated category are shown in Table~\ref{table:SE+SS}. The proposed DESNet gives the best performance in all three tracks. In SE track, DCCRN is the second only to DESNet, showing its powerful capability of noise reduction. Conv-TasNet is even worse than CACGMM probably due to the way it utilizes the spatial information is limited and not effective as FasNet. Staged SNR has less impact on SE task while beam feature improves PESQ by 0.21. According to the results in CSS/NSS track, DPRNN and FasNet are superior to DCCRN and Conv-TasNet, achieving around 3 dB improvement on SI-SNR in both test sets. The staged SNR strategy and beam feature are important for introducing better separation results, bringing 1.22/1.36 dB and 0.88/1.39 dB improvement in terms of SI-SNR in CSS and NSS tracks, respectively. Symphonic loss brings 0.13 gain on PESQ, as well as 0.56 and 0.44 dB gain on SI-SNR in CSS and NSS tracks, respectively. Compared with FasNet, the best model in our baseline system, DESNet achieves 0.72/0.23 dB improvements on CSS/NSS tracks.

The evaluation results of dereverberated category are shown in Table~\ref{table:SE+SS+DRV}. Removing WPE module has much impact on separation performance, leading the degradation of 0.59/1.00 dB on CSS/NSS tracks in terms of SI-SNR. DESNet outperforms WPE-DCCRN baseline and brings 1.96 dB and 1.56 dB improvements on those two tracks. Staged SNR and symphonic loss are proved again to be effective, bringing 0.09/0.48 dB and 0.45/0.30 dB improvement in terms of SI-SNR in CSS and NSS tracks, respectively.

In order to see what the attentional selection layer learnt in the DESNet, we split the whole utterance into consequent chunks to get the time-variant weights and visualize an example of a two-speaker mixture utterance on angle feature in Fig.~\ref{fig:weight}. The $x$ axis denotes chunk index and $y$ axis denotes the direction of the speaker. The real direction of two speakers are $102^\circ$ and $193^\circ$ and we can see that the learnt weights fit the actual speaker's direction perfectly, which can select the correct angle feature that we want the network to learn. Audio samples are available online\footnote{\url{https://felixfuyihui.github.io/DesNet_Demo/}}.

\section{Conclusions}
In this paper, we propose a multi-channel network named DESNet for simultaneous speech dereverberation, enhancement and separation. The novel DCCRN is utilized to perform speech unmixing and form the pre-unmixing masks to assist the soft selection of angle and beam features used in subsequent speech extraction network. The neural network based WPE is cascaded to produce dereverberated signal. All modules mentioned above are jointly optimized to form an end-to-end manner. To lead better performance in each task, we also introduce \textit{staged SNR} strategy and \textit{symphonic loss}. Experiments show that the proposed DESNet outperforms DCCRN in enhancement task and FasNet in separation task on non-dereverberated cases. In dereverberated category, DESNet also shows improvements over the cascaded WPE-DCCRN network in enhancement and separation tasks. In the future, we will consider optimizing speech dereverberation, enhancement and separation with acoustic model to further improve the speech recognition accuracy in real environment scenarios.

\clearpage
\balance
\bibliographystyle{IEEEbib}
\bibliography{refs}

\begin{thebibliography}{10}

\bibitem{hershey2016deep}
John~R Hershey, Zhuo Chen, Jonathan Le~Roux, and Shinji Watanabe,
\newblock ``Deep clustering: Discriminative embeddings for segmentation and
  separation,''
\newblock in {\em 2016 IEEE International Conference on Acoustics, Speech and
  Signal Processing (ICASSP)}. IEEE, 2016, pp. 31--35.

\bibitem{valentini2016investigating}
Cassia Valentini-Botinhao, Xin Wang, Shinji Takaki, and Junichi Yamagishi,
\newblock ``Investigating {RNN}-based speech enhancement methods for
  noise-robust text-to-speech.,''
\newblock in {\em SSW}, 2016, pp. 146--152.

\bibitem{kinoshita2016summary}
Keisuke Kinoshita, Marc Delcroix, Sharon Gannot, Emanu{\"e}l~AP Habets,
  Reinhold Haeb-Umbach, Walter Kellermann, Volker Leutnant, Roland Maas,
  Tomohiro Nakatani, Bhiksha Raj, et~al.,
\newblock ``A summary of the reverb challenge: state-of-the-art and remaining
  challenges in reverberant speech processing research,''
\newblock {\em EURASIP Journal on Advances in Signal Processing}, vol. 2016,
  no. 1, pp. 7, 2016.

\bibitem{tan2018gated}
Ke~Tan, Jitong Chen, and DeLiang Wang,
\newblock ``Gated residual networks with dilated convolutions for monaural
  speech enhancement,''
\newblock {\em IEEE/ACM transactions on audio, speech, and language
  processing}, vol. 27, no. 1, pp. 189--198, 2018.

\bibitem{yin2020phasen}
Dacheng Yin, Chong Luo, Zhiwei Xiong, and Wenjun Zeng,
\newblock ``Phasen: A phase-and-harmonics-aware speech enhancement network.,''
\newblock in {\em AAAI}, 2020, pp. 9458--9465.

\bibitem{hu2020dccrn}
Yanxin Hu, Yun Liu, Shubo Lv, Mengtao Xing, Shimin Zhang, Yihui Fu, Jian Wu,
  Bihong Zhang, and Lei Xie,
\newblock ``{DCCRN}: Deep complex convolution recurrent network for phase-aware
  speech enhancement,''
\newblock {\em arXiv preprint arXiv:2008.00264}, 2020.

\bibitem{luo2019conv}
Yi~Luo and Nima Mesgarani,
\newblock ``Conv-tasnet: Surpassing ideal time--frequency magnitude masking for
  speech separation,''
\newblock {\em IEEE/ACM transactions on audio, speech, and language
  processing}, vol. 27, no. 8, pp. 1256--1266, 2019.

\bibitem{luo2020dual}
Yi~Luo, Zhuo Chen, and Takuya Yoshioka,
\newblock ``Dual-path {RNN}: efficient long sequence modeling for time-domain
  single-channel speech separation,''
\newblock in {\em ICASSP 2020-2020 IEEE International Conference on Acoustics,
  Speech and Signal Processing (ICASSP)}. IEEE, 2020, pp. 46--50.

\bibitem{liu2019divide}
Yuzhou Liu and DeLiang Wang,
\newblock ``Divide and conquer: A deep casa approach to talker-independent
  monaural speaker separation,''
\newblock {\em IEEE/ACM Transactions on Audio, Speech, and Language
  Processing}, vol. 27, no. 12, pp. 2092--2102, 2019.

\bibitem{zeghidour2020wavesplit}
Neil Zeghidour and David Grangier,
\newblock ``Wavesplit: End-to-end speech separation by speaker clustering,''
\newblock {\em arXiv preprint arXiv:2002.08933}, 2020.

\bibitem{takahashi2019recursive}
Naoya Takahashi, Sudarsanam Parthasaarathy, Nabarun Goswami, and Yuki
  Mitsufuji,
\newblock ``Recursive speech separation for unknown number of speakers,''
\newblock {\em arXiv preprint arXiv:1904.03065}, 2019.

\bibitem{wichern2019wham}
Gordon Wichern, Joe Antognini, Michael Flynn, Licheng~Richard Zhu, Emmett
  McQuinn, Dwight Crow, Ethan Manilow, and Jonathan~Le Roux,
\newblock ``{WHAM}!: Extending speech separation to noisy environments,''
\newblock {\em arXiv preprint arXiv:1907.01160}, 2019.

\bibitem{maciejewski2020whamr}
Matthew Maciejewski, Gordon Wichern, Emmett McQuinn, and Jonathan Le~Roux,
\newblock ``{WHAMR}!: Noisy and reverberant single-channel speech separation,''
\newblock in {\em ICASSP 2020-2020 IEEE International Conference on Acoustics,
  Speech and Signal Processing (ICASSP)}. IEEE, 2020, pp. 696--700.

\bibitem{ma2020two}
Chao Ma, Dongmei Li, and Xupeng Jia,
\newblock ``Two-stage model and optimal si-snr for monaural multi-speaker
  speech separation in noisy environment,''
\newblock {\em arXiv preprint arXiv:2004.06332}, 2020.

\bibitem{wang2018integrating}
Zhong-Qiu Wang and DeLiang Wang,
\newblock ``Integrating spectral and spatial features for multi-channel speaker
  separation,''
\newblock in {\em Interspeech}, 2018, pp. 2718--2722.

\bibitem{yoshioka2019low}
Takuya Yoshioka, Zhuo Chen, Changliang Liu, Xiong Xiao, Hakan Erdogan, and
  Dimitrios Dimitriadis,
\newblock ``Low-latency speaker-independent continuous speech separation,''
\newblock in {\em ICASSP 2019-2019 IEEE International Conference on Acoustics,
  Speech and Signal Processing (ICASSP)}. IEEE, 2019, pp. 6980--6984.

\bibitem{wu2020end}
Jian Wu, Zhuo Chen, Jinyu Li, Takuya Yoshioka, Zhili Tan, Ed~Lin, Yi~Luo, and
  Lei Xie,
\newblock ``An end-to-end architecture of online multi-channel speech
  separation,''
\newblock {\em arXiv preprint arXiv:2009.03141}, 2020.

\bibitem{wu2020saddel}
Yuan-Kuei Wu, Chao-I Tuan, Hung-yi Lee, and Yu~Tsao,
\newblock ``{SADDEL}: Joint speech separation and denoising model based on
  multitask learning,''
\newblock {\em arXiv preprint arXiv:2005.09966}, 2020.

\bibitem{kinoshita2018listening}
Keisuke Kinoshita, Lukas Drude, Marc Delcroix, and Tomohiro Nakatani,
\newblock ``Listening to each speaker one by one with recurrent selective
  hearing networks,''
\newblock in {\em 2018 IEEE International Conference on Acoustics, Speech and
  Signal Processing (ICASSP)}. IEEE, 2018, pp. 5064--5068.

\bibitem{von2019all}
Thilo von Neumann, Keisuke Kinoshita, Marc Delcroix, Shoko Araki, Tomohiro
  Nakatani, and Reinhold Haeb-Umbach,
\newblock ``All-neural online source separation, counting, and diarization for
  meeting analysis,''
\newblock in {\em ICASSP 2019-2019 IEEE International Conference on Acoustics,
  Speech and Signal Processing (ICASSP)}. IEEE, 2019, pp. 91--95.

\bibitem{kinoshita2020tackling}
Keisuke Kinoshita, Marc Delcroix, Shoko Araki, and Tomohiro Nakatani,
\newblock ``Tackling real noisy reverberant meetings with all-neural source
  separation, counting, and diarization system,''
\newblock in {\em ICASSP 2020-2020 IEEE International Conference on Acoustics,
  Speech and Signal Processing (ICASSP)}. IEEE, 2020, pp. 381--385.

\bibitem{nakatani2020dnn}
Tomohiro Nakatani, Riki Takahashi, Tsubasa Ochiai, Keisuke Kinoshita, Rintaro
  Ikeshita, Marc Delcroix, and Shoko Araki,
\newblock ``{DNN}-supported mask-based convolutional beamforming for
  simultaneous denoising, dereverberation, and source separation,''
\newblock in {\em ICASSP 2020-2020 IEEE International Conference on Acoustics,
  Speech and Signal Processing (ICASSP)}. IEEE, 2020, pp. 6399--6403.

\bibitem{kinoshita2017neural}
Keisuke Kinoshita, Marc Delcroix, Haeyong Kwon, Takuma Mori, and Tomohiro
  Nakatani,
\newblock ``Neural network-based spectrum estimation for online {WPE}
  dereverberation.,''
\newblock in {\em Interspeech}, 2017, pp. 384--388.

\bibitem{reddy2020interspeech}
Chandan~KA Reddy, Ebrahim Beyrami, Harishchandra Dubey, Vishak Gopal, Roger
  Cheng, Ross Cutler, Sergiy Matusevych, Robert Aichner, Ashkan Aazami,
  Sebastian Braun, et~al.,
\newblock ``The {I}nterspeech 2020 deep noise suppression challenge: Datasets,
  subjective speech quality and testing framework,''
\newblock {\em arXiv preprint arXiv:2001.08662}, 2020.

\bibitem{Luo2019End}
Yi~Luo, Zhuo Chen, Nima Mesgarani, and Takuya Yoshioka,
\newblock ``End-to-end microphone permutation and number invariant
  multi-channel speech separation,''
\newblock in {\em ICASSP 2020-2020 IEEE International Conference on Acoustics,
  Speech and Signal Processing (ICASSP)}. IEEE, 2020, pp. 6394--6398.

\bibitem{le2019sdr}
Jonathan Le~Roux, Scott Wisdom, Hakan Erdogan, and John~R Hershey,
\newblock ``{SDR}--half-baked or well done?,''
\newblock in {\em ICASSP 2019-2019 IEEE International Conference on Acoustics,
  Speech and Signal Processing (ICASSP)}. IEEE, 2019, pp. 626--630.

\bibitem{ito2016complex}
Nobutaka Ito, Shoko Araki, and Tomohiro Nakatani,
\newblock ``Complex angular central gaussian mixture model for directional
  statistics in mask-based microphone array signal processing,''
\newblock in {\em 2016 24th European Signal Processing Conference (EUSIPCO)}.
  IEEE, 2016, pp. 1153--1157.

\end{thebibliography}

\end{document}